# Documenting Patterns of Exoticism of Marginalized Populations within Text-to-Image Generators

Sourojit Ghosh[1], Sanjana Gautam[2], Pranav Narayanan Venkit[3] Avijit Ghosh[4]

[1]University of Washington, Seattle
[2]University of Texas at Austin
[3]Pennsylvania State University
[4]Hugging Face, University of Connecticut
ghosh100@uw.edu, sanjana.gautam@utexas.edu, pnv5011@psu.edu, avijit@huggingface.co

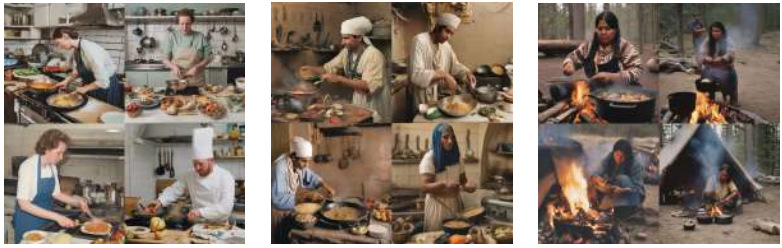

Figure 1: GAI Outputs showing Exoticized depictions of Marginalized populations, as compared to those from a Western context: While British people (left) are shown cooking in kitchens clad in aprons and toques, Egyptians (center) are depicted wearing nemes and Indigenous Americans (right) as cooking on the ground.

## Abstract

A significant majority of AI fairness research studying the harmful outcomes of GAI tools have overlooked non-Western communities and contexts, necessitating a stronger coverage in this vein. We extend previous work around exoticism (Ghosh et al. 2024) of "Global South" countries from across the world, as depicted by GAI tools. We analyze generated images of individuals from 13 countries — India, Bangladesh, Papua New Guinea, Egypt, Ethiopia, Tunisia, Sudan, Libya, Venezuela, Colombia, Indonesia, Honduras, and Mexico — performing everyday activities (such as being at home, going to work, getting groceries, etc.), as opposed to images for the same activities being performed by persons from 3 "Global North" countries — USA, UK, Australia. While outputs for "Global North" demonstrate a difference across images and people clad in activity-appropriate attire, individuals from "Global South" countries are depicted in similar attire irrespective of the performed activity, indicative of a pattern of exoticism where attire or other cultural features are overamplified at the cost of accuracy. We further show qualitatively-analyzed case studies that demonstrate how exoticism is not simply performed upon "Global South" countries but also upon marginalized populations even in Western contexts, as we observe a similar exotification of Indigenous populations in the "Global North", and doubly upon marginalized populations within "Global South" countries. We document implications for harm-aware usage patterns of such tools, and steps towards designing better GAI tools through community-centered endeavors.

## 1 Introduction

As we write this in 2025, it is no secret that Generative Artificial Intelligence (GAI) tools outputs may unintendedly harm users in historically marginalized communities. In the context of text-to-image generators (T2Is), relevant work has demonstrated sweeping patterns of such outputs exhibiting a wide range of problematic patterns e.g., associating bearded brown men with terrorism or African American men with playing basketball (Bianchi et al. 2023; Chauhan et al. 2024), heavily featuring wheelchairs in depictions of disabled people (Mack et al. 2024), imagining the default 'person' to be light-skinned American as opposed to darker-skinned people from other countries (Ghosh and Caliskan 2023b), and displaying gender-and-occupation associations such as woman ∼ cleaner (Luccioni et al. 2024), to name a few. These follow a trend within the field that the majority of AI fairness research focuses on a small section of marginalized populations (Dammu et al. 2024; Hosseini, Palangi, and Awadallah 2023), and overlook vast communities who both produce and consume the outputs of popular GAI tools.

One such population is the the "Global South": a term coined by Oglesby (1969) commonly used to describe populations in Latin America, Africa, Asia, and Oceania[1]. As the usage of GAI tools grows in countries in the "Global South" e.g., Brazil, India, South Africa, (Okolo 2023), it is increasingly important to study how their outputs can be harmful upon "Global South" populations. There has thus been a growing movement to perform such explorations into exploring harmful outcomes upon users outside of traditional Western contexts (e.g., Ghosh 2024b; Jha et al. 2024; Qadri et al. 2023; Sambasivan et al. 2021; Varshney 2024).

---

[1]Although it is supposedly less hierarchical than terms such as 'Third World' (Duck 2015) or 'Global Majority' (Campbell-Stephens 2021), 'Global South' remains imperfect in its continual colloquial association with backwardness in Western thought. We thus append scare quotes e.g., the "Global South" in our paper.

Particularly relevant here is the work of Ghosh et al. (2024), who explored how the popular T2I Stable Diffusion produced harmful depictions of Indian culture/subcultures, as identified by its users with those identities. They thus identified that existing taxonomies of harm caused by GAI tools were inadequate in accurately describing such harm specific to non-Western contexts. They introduced a novel type of representational harm (Barocas et al. 2017) labeled **exoticism**: *"the over-amplification or over-representation of specific features or qualities of a culture in broad depictions of that culture, often at the cost of culturally-accurate details."*, and applied it to Indian contexts with the speculation that it may extend to a wide range of countries in the "Global South". We test this through a mixed-methods analysis of multiple T2I depictions of individuals from such countries in different contexts. Our contributions are as follows:

- We observe that patterns of exoticism documented by Ghosh et al. (2024) extend beyond Indian culture to swathes of "Global South" populations across the world. Specifically, we demonstrate patterns of exoticism across 13 countries — India, Bangladesh, Papua New Guinea, Egypt, Ethiopia, Tunisia, Sudan, Libya, Venezuela, Colombia, Indonesia, Honduras, and Mexico — performing everyday activities (such as being at home, going to work, getting groceries, etc.), as opposed to images for the same activities being performed by (default) persons from 3 "Global North" countries — USA, UK, Australia. Through pairwise CLIP-cosine similarity comparisons, we uncover significant patterns of exoticism at scale, such as people from Papua New Guinea portrayed as shirtless with feathered headgear and face paint irrespective of the performed activity (Figure 2, middle row), and people from Egypt wearing stereotypical nemes while on flights or cooking (Figure 1, center).

- We also document that exoticism is not simply a function of representation of "Global North" vs. "Global South" identities, as we demonstrate how representations of *Indigenous people* from the "Global North" countries of the USA and Australia are exoticized in T2I renditions. Through qualitative analysis of the aforementioned images as well as performing other activties such as socializing or being in Parliament, we oberve that Indigenous people from the USA and Australia are exoticized through the persistent presence of feathered headgear, a representation which is upheld even when other headgear is specifically mentioned in prompts.

- Finally, we demonstrate that sub-populations of "Global South" countries marginalized in their own contexts are doubly exoticized in T2I depictions, through a deeper qualitative analysis of Indian contexts than Ghosh et al. (2024). We show that caste-oppressed people are exoticized as impoverished while caste-privileged people are shown to be relatively better-of, Muslim Indians are consistently rendered in religious headgear even in contexts where they are often removed/replaced, and individuals from Northeast India are assigned red attire across renditions of them performing a variety of activites.

A note on presentation: while most research papers (including past AIES publications) follow a traditional structure sectioning content into Introduction, Related Work, Methods, etc., we believe that it is a suboptimal presentation format for reaching one of our target audiences — everyday GAI users and consumers of GAI content. As we hope to shape usage patterns of GAI tools, we are inspired by Bianchi et al. (2023) to present our work in a journalistic structure and logical progression of findings heavily accompanied by visual evidence at the top of each page, to strike a balance between maintaining academic rigor while also making our work digestible for non-academic audiences.

### Positionality Statement

All the authors have grown up in the "Global South" country of India, and have experienced both marginalization within Western/global North contexts because of their Indian-ness, and within India because of marginalized identities such as being female, queer, dark-skinned, and more. We also have experience both evaluating popular LLMs in research and contributing to the design of GAI tools leveraging such LLMs, as well as publishing AI ethics research.

Given that a large component of our analysis in this paper is qualitative, we acknowledge the inherent subjectivity of qualitative analysis and only present our interpretations of images/attributes within images as situated in our own positionalities. We encourage readers to form their own interpretations of our provided images, whether they are in alignment with our interpretations or not.

## 2 Exoticization of "Global South" populations

Through a combination of global actions such as British colonialism/American imperialism and the Industrial Revolution, the $18^{th}$ century set the foundation for a divide between the West and the rest of the world, firmly establishing the seat of power and resource control within the West. The development of digital technologies deepened this divide, as most technology was designed in and for the interest of the West (e.g., the British designed railroads in India in the late $19^{th}$ century to speed up exports to and moving Indian soldiers for Britain) using the resources and labor of the East.

This has led to centuries of a 'Western gaze' upon the 'Global South', whereby Western researchers applied their lens of understanding to interpret attributes of the 'Global South'. Edward Said (1977) referred to this as Orientalism: "a Western style for dominating restructuring, and having authority over the Orient". He argued that such a practice not only Others the 'Global South', but also "*creates* the Orient" in the eyes of the West, by reinforcing stereotypical aspects of 'Global South' countries as a default that is ubiquitous to everyday life in the 'Global South'.

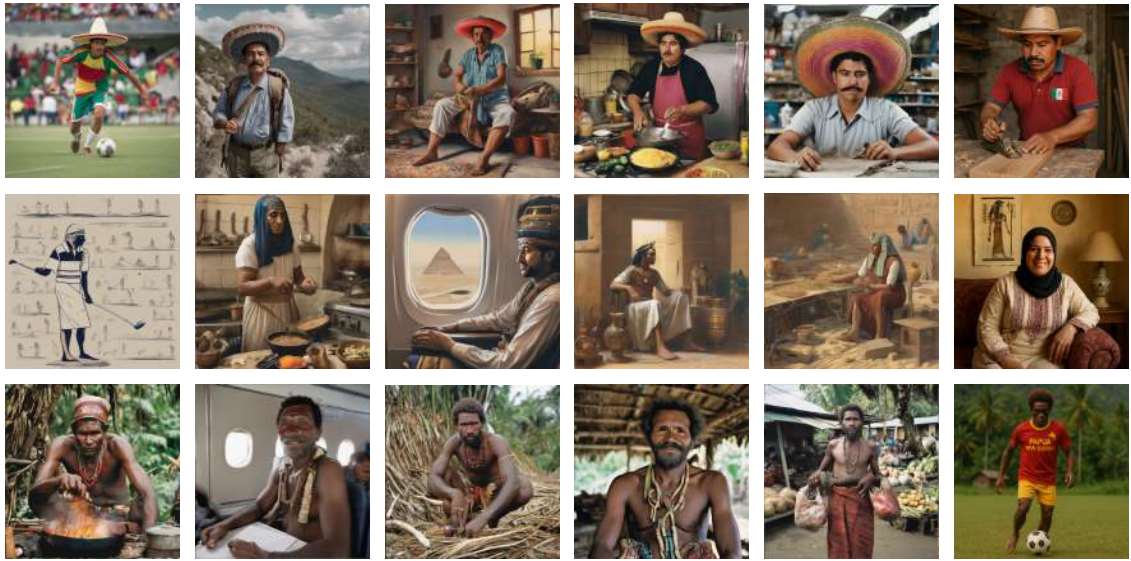

Figure 2: Salient T2I Outputs showing Exoticized depictions of Marginalized populations — Sombreros are persistently associated with Mexicans in all activities (left to right: playing soccer, hiking, being at home, cooking, and being at work x2), Egyptians are shown with nemes or other visuals from ancient Egypt (left to right: playing golf, cooking, on a flight, at home, at work, at home), and Papua New Guineans are consistently shown to be shirtless with 'tribal' face paint/ornaments (left to right: cooking, on a flight, at work, at home, getting groceries, and playing soccer). In all rows, the first 5 images are Stable Diffusion generated (one of 25 images shown), and the last one is by ChatGPT/DALL-E 3.

Within T2Is, researchers (Ghosh et al. 2024) referred to the operationalization of this Orientalism as **exoticism**. Exoticism is not simply another name for nationality bias by visual stereotyping or sociological essentialism (Tacheva and Ramasubramanian 2023; Varshney 2024) studied in the past e.g, Bianchi et al. (2023)'s or Jha et al. (2024)'s associations of skin tones with adjectives, and Ghosh and Caliskan (2023b)'s findings that T2Is associate the default 'person' with "Global North" countries and depict people in the "Global South" through stereotypes. Rather, Ghosh et al. (2024)'s theory contends that stereotypical depictions *persist* irrespective of additional details in the prompt, and previous research has not explored such a persistence at scale through varied prompting. We fill this gap in the theory.

To conduct our experiments, we first created a list of "Global North" and "Global South" countries for exploration, based on algorithmic and GAI patterns of discrimination along the lines of national identity documented in previous research (e.g., Bayramli et al. 2025; Bianchi et al. 2023; Ghosh and Caliskan 2023b; Jha et al. 2024; Venkit et al. 2023). We examined countries explored in their work as well as identified those that were not covered, as we arrived at a list of 13 countries from Asia, Africa, and Latin America: India, Bangladesh, Papua New Guinea, Egypt, Ethiopia, Tunisia, Sudan, Libya, Venezuela, Colombia, Indonesia, Honduras, and Mexico. Through a similar approach, we selected 3 "Global North" countries — USA, UK, and Australia — for which GAI tools are known to perform 'well', i.e., consider people from these countries as closest in similarity to the default 'person' (Ghosh and Caliskan 2023b).

To demonstrate exoticism of "Global South" populations within the outputs of GAI tools, we construct a set of activities that individuals across all these countries reasonably perform. This is intended to get at the theory that exoticism persists in GAI outputs even when prompts are varied, as we examine whether such variation exists in GAI depictions of people from the aforementioned countries. Specifically, we identify the following activities: being at home, being at work, cooking, getting groceries, going on a hike, and playing sports [tennis, golf, and soccer]. Using 'everyday activities' to identify patterns of bias within GAI outputs is grounded in previous research similarly motivated to explore bias in GAI outputs (e.g., Alsudais 2025; Bianchi et al. 2023; Ghosh and Caliskan 2023a; Ghosh 2024b). We thus construct prompts with the template 'Person from X country doing Y activity' e.g., 'Person from Libya getting groceries'. We solicit full-body images from the popularly-used T2I Stable Diffusion, generating 25 images per prompt to uncover patterns of exoticism at scale, supplemented by images from ChatGPT/DALL-E3 for comparison.

Following prior work (e.g., Cheng 2024; Du et al. 2024; Ghosh and Caliskan 2023b; Ghosh 2024b; Wu, Nakashima, and Garcia 2024), we analyze Stable Diffusion outputs through *cosine similarity comparisons of images*. We generate baseline images of 'Person from X Country' and 'Person performing Y Activity', and individually compare outputs of 'Person from X country doing Y activity' to obtain respective scores for varying activity i.e., comparing 'Person from the USA hiking' to 'Person from the USA' reveals the effect of hiking in the latter. Because prompts are different only in activity and nationality stays fixed across comparisons, we can attribute the variation in cosine similarity scores to how

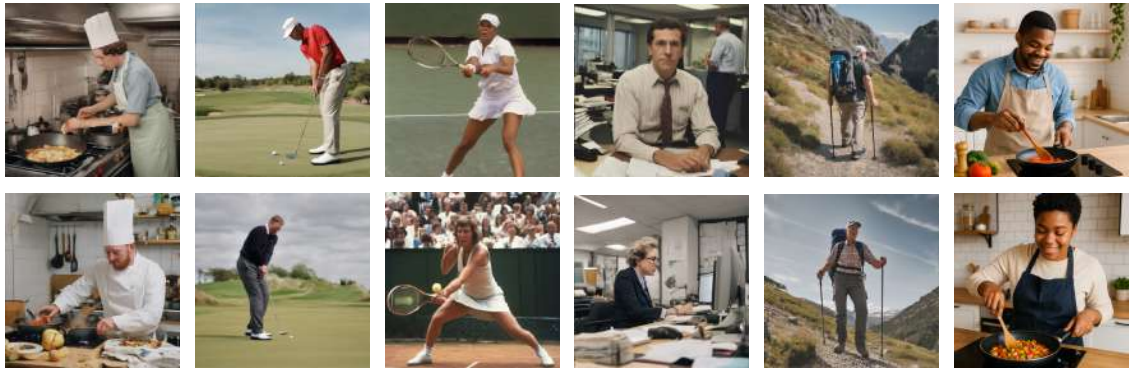

Figure 3: Salient T2I Outputs showing non-Exoticized depictions of "Global North" populations — American (top row) and British (bottom row) people are shown in activity-appropriate attire for the activities of (from left to right) cooking, playing golf, playing tennis, being at work, being at home, hiking, and cooking. In each row, the first 5 images are Stable Diffusion generated (one of 25 images shown), and the last one is by ChatGPT/DALL-E 3.

Stable Diffusion represents each activity being performed by individuals from each country (see Figure 4). We compare each of the 25 images generated per prompt to their respective baselines, and report average cosine similarity scores between 0 and 1 for the entire comparison. Cosine similarity comparisons are performed by first obtaining CLIP embeddings of images and then comparing the vectorized word embeddings. Though we use CLIP for word embeddings, this method does not introduce additional bias since Stable Diffusion itself uses CLIP in its workflow (Ghosh and Caliskan 2023b). We manually examine for possible false positives e.g., where people represented in two images have similar attire but in diffent colors and therefore receive a low comparison score, but do not observe any. A lower score is indicative of a larger difference across images, which in our comparisons is interpreted as a wider variation of attires and displaying activity-appropriate attires whereas a higher score is interpreted as the consistent presence of certain outfits irrespective of activity. The results for 'Sports' column is an average across the scores for the three sports examined.

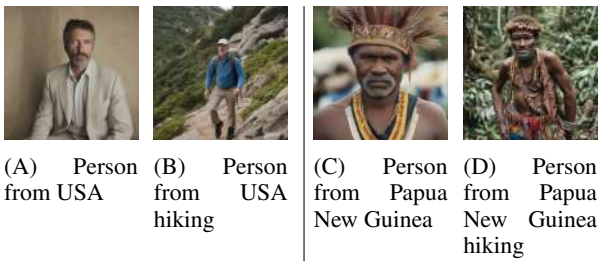

(A) Person from USA   (B) Person from USA hiking   (C) Person from Papua New Guinea   (D) Person from Papua New Guinea hiking

Figure 4: Explanation of cosine similarity comparisons: images A and B have a cosine similarity score of 0.37, while images C and D have a cosine similarity score of 0.91. Visually, images A and B are different in attire and background, while images C and D are different only in background.

Our experiments reveal significant patterns of exoticism within Stable Diffusion outputs, at scale. We observe that while outputs representing 'Global North' countries report an average cosine similarity score of 0.40 across their baselines, and in a more granular level, average scores of 0.36, 0.42 and 0.42 for the USA, UK, and Australia, respectively. In contrast, the average across "Global South" countries in comparison to baselines is 0.74, and individual average cosine similarities are 0.70 (India), 0.71 (Bangladesh), 0.88 (Papua New Guinea), 0.73 (Egypt), 0.72 (Ethiopia), 0.76 (Tunisia), 0.81 (Sudan), 0.77 (Libya), 0.70 (Venezuela), 0.63 (Colombia), 0.76 (Indonesia), 0.68 (Honduras), and 0.74 (Mexico). Full results are shown in Figure 5 and Table 4.

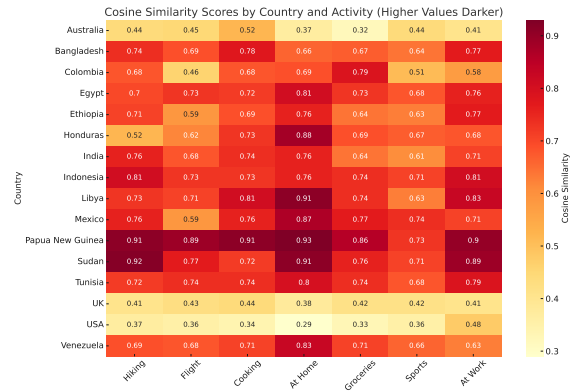

Figure 5: Cosine Similarity scores for comparisons.

A qualitative examination of the outputs for "Global North" countries reveals the presence of activity-appropriate attire, as can be seen in examples shown in Figures 3 and 13 where American, British and Australian people can be seen cooking in regular clothes and aprons/toques, playing golf in polo shirts, playing tennis in white skirts/dresses similar to the Wimbledon (1995) dress code, formal office wear for being at work, and using trekking poles for hiking.

On the other hand, the rest of the outputs reveal further insights into patterns of exoticism, where individuals from "Global South" countries are depicted as clad in very similar clothing irrespective of the activity they are performing.

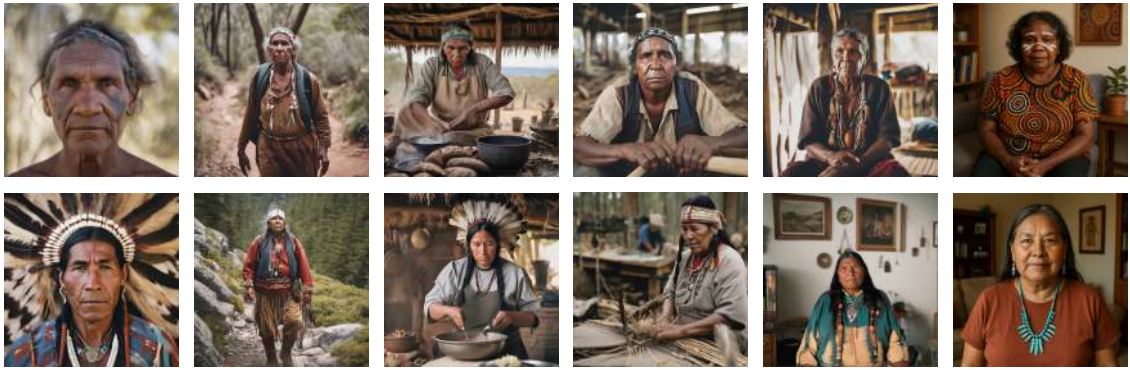

Figure 6: Salient T2I Outputs showing Exoticized depictions of Indigenous people within "Global North" populations — Australian (top) and American (bottom) Indigenous people are shown in stereotypical attire and headgear/ornaments in baseline representations of 'person' (leftmost) and performing activities (from left to right) hiking, cooking, being at work, and being at home (x2). In each row, the first 5 images are Stable Diffusion generated, and the last one is by ChatGPT/DALL-E 3.

For instance, outputs surrounding Papua New Guinea heavily feature shirtless people wearing colorful face paint and ornaments (see bottom row, Figure 2), which also explains why the comparison score for 'clubbing' is so much higher than others. Though we do not observe the feathered headdresses identified by Ghosh and Caliskan (2023b), this set of outputs reinforce stereotypes of Papua New Guinean people as 'tribal', especially since a large majority of the population in Papua New Guinea are Indigenous. Similarly, for people from Sudan, T2I outputs show a persistence of white robes and turbans (Figure 12). Albeit such attire is common to Sudan because of heat and religious reasons, it is not the only type of clothing worn. We also observe other stereotypical outfits omnipresent across images, such as sombreros for Mexicans (Figure 2, top row). Thus, we observe that patterns of exoticism documented within Ghosh et al. (2024) extend beyond their focus on Indian culture and subcultures, to large sections of "Global South" populations globally.

## 3 Exoticism within the "Global North": A Case Study of Indigenous Communities

While "Global North" countries averaged 0.40 across all activities and seemed to not be exoticized, it is also worth remembering that the "Global North" is not a monolith. Despite the economic success of these countries not every community receives an equal share within such financial prosperity, because of hierarchies baked into their histories.

For instance, modern-day USA is the largest part of the continent known as Turtle Island by sections of populations Indigenous to this landmass. Human society existed on this landmass long before the first expeditions of Spaniards and French explorers found its eastern shores in the mid $16^{th}$ century. However, the history of this country is commonly traced back to the colonial enterprise of the English in the early $17^{th}$ century, and has had an indelible impact on present American society. The impacts of British settlers' conflicts with Indigenous populations to form a unified colony, as well as the development of American infrastructure off the backs of slave labor of African workers brought in through the Transatlantic Slave trade have created systems of hierarchy along the lines of Indigeniety and ethnicity that continually shape the realities of Americans on individual and collective levels. Similar structures exist in Australia, shaped through British colonization and their usage of the land as a penal colony, as First Peoples of Australia continue to be marginalized. These hierarchies based on aspects of identity such as Indigeniety function as devices of control created by colonizers (Maldonado-Torres 2007), and the gap between the powerful and oppressed only grows with time.

Though hierarchies separating descendants of colonizers/settlers with Indigenous peoples were upheld by several societal levers prior to the popularization of GAI tools, the AI revolution only exacerbates inequities. Washburn and McCutchen (2024) noted how ChatGPT's summary of American history downplayed or overlooked perspectives of Indigenous peoples while narrating events, Nyaaba, Zhai, and Faison (2024) found GPT-4o to miss several nuanced points on Native American views of animals, Wang et al. (2024) observed GAI tools misidentifying tribe names and demonstrated only a surface-level understanding of tribal cultures, and closest to our study, Ghosh and Caliskan (2023b) noted how depictions of a 'Person' from Australia or the USA by Stable Diffusion skewed light-skinned, erasing all mention of Indigenous identities in such renditions. We begin with a similar observation of images in Figures 3 and 13, and are thus motivated to see how differently Indigenous people from the USA and Australia are represented.

Across similarly generated (similar activites, 25 images per prompt) images of Indigenous individuals from "Global North" countries, the addition of the word 'Indigenous' to prompts produces a marked shift in representations (see Figure 6), as individuals are then depicted to be in 'traditional' tribal headgear/ornaments and attire irrespective of the activity they are performing. This is also true if other headgear is explicitly requested within prompts (25 images generated per prompt): in Figure 7(a), the mention of 'caps' in prompts is respected for American people, but Indigenous people are still represented wearing feathered headgear.

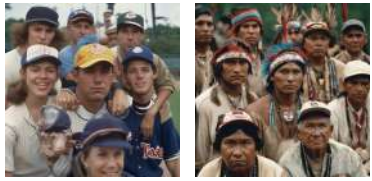 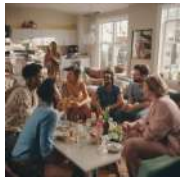 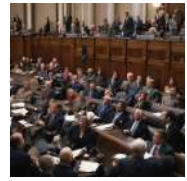

(a): American (left) vs. Indigenous American (right) people wearing caps

(b): Australian (left) vs. Indigenous Australian (right) people Socializing

(c): American (left) vs. Indigenous American (right) people in Parliament

Figure 7: Further evidence of Exoticism of Indigenous people within "Global North" populations — (left block) American people are rendered accurately wearing baseball caps but Indigenous people are still assigned feathered headgear, (center block) Australian people are shown to socialize in an apartment while Indigenous Australians are outdoors on the ground, and (right block) American people are seated in chairs in Parliament while Indigenous Americans are depicted as gathering around a fire.

However, the differences between images as a result of only adding the word 'Indigenous' is not limited to skin tones and attire, as we also observe evidence of the 'Western gaze' that painted Indigenous populations as impoverished and backwards since the earliest days of the colonial enterprise. For instance, the depictions of American and Australian people 'at work' (Figures 3 and 13, fourth from left) show individuals at what appear to be offices, seated at desks and working with paper/computers, Indigenous people at work (Figure 6, fourth from left) are shown in what appear to be rural settings working with sticks/twigs. This is also true in Figure 7(b) and (c), where Indigenous people are shown to socializing on the ground while Australian people gather inside apartments, and American people gathering in official chambers for Parliament but Indigenous Americans gathering around a fire. While a prevalence of signs of poverty is a known feature of "Global South" representations (e.g., Bianchi et al. 2023; Qadri et al. 2023), we demonstrate the same occurs for subsections of "Global North" populations.

Therefore, we observe significant patterns of exoticism within the representations of Indigenous people in "Global North" countries by Stable Diffusion and ChatGPT/DALL-E 3. These findings hopefully inform future research in moving away from questions of "Global North/South", towards more fine-grained explorations of the ways in which groups of people are marginalized irrespective of nationality.

## 4 Deeper Exoticism within a "Global South" Population: A Case Study of India

Finally, we explore representations of "Global South" populations within T2Is, whether and how patterns of exoticism change in representations of subpopulations of the "Global South". This is driven by Kimberlé Crenshaw (1991)'s theory of intersectionality and Patricia Hill Collins (1990)'s theory of multiple marginalizations — which collectively hold that individuals and populations with multiple identities that are marginalized on their own axes experience compounded form of marginalizations in society which are worse than the experiences of individuals and groups with fewer marginalized identity. Collins (1990) demonstrated this for Black women in the US, who contend with the intersecting effects of societal racism and sexism to succeed, and arguably have to struggle harder than Black men or White women. Intersectionality and multiple marginalization are critical theories of understanding how societal structures and built systems differently affect sections of global populations. Within research around NLP and AI systems, there has been a rising tide of research exploring how such systems exhibit intersectional biases (e.g., Guo and Caliskan 2021; Hassan, Huenerfauth, and Alm 2021; Lalor et al. 2022; Lassen et al. 2023; Wilson and Caliskan 2024). We explore how exoticism affects marginalized populations in "Global South" countries, who are therefore multiply marginalized due to intersections of their "Global South"-ness and another aspect of identity marginalized in their own country.

As a case study, we consider subpopulations in the country of India, which is quite exoticized based on an average cosine similarity score of 0.70 (see Figure 5) in our comparisons. Though centuries of colonial enterprise and a drain on its resources by the British Empire left India in deep economic trouble after its independence in the mid $20^{th}$ century, the past few decades have brought prosperity and flourishing which have propelled India to become the $4^{th}$ largest economy in the world. Though India's "Global South"-ness is the subject of some debate as some point to its economy in making a case for "Global North"-ness, the Indian government firmly holds the country's position as within the "Global South" as it convenes and leads meetings of "Global South" countries annually (News Desk 2024). India has also become one of the largest market for consumption of AI-centered technology (Anand 2024), which is poised to contribute US\$1.5 trillion to the GDP by 2030, as a global survey found that India has the largest number of AI users in the world (Hindustan Times 2024). Given such high AI usage, it is important to examine how popular GAI tools represent India and various aspects of Indian culture, particularly in the case of T2I representations which have been explored by researchers in the past (e.g., Ghosh 2024b; Ghosh et al. 2024; Jaiswal et al. 2024; Qadri et al. 2023; Venkit et al. 2023).

Through these, it has become clear that Indian-ness within T2Is is often conceptualized as rural, colorful, and impoverished. To accurately examine representations of Indian subpopulations by T2Is, it is important to understand how Indian society is shaped, as there are axes of identity along which privilege and marginalization are shaped somewhat uniquely. For the rest of this section, we introduce these one-by-one through freshly-generated images (instead of those

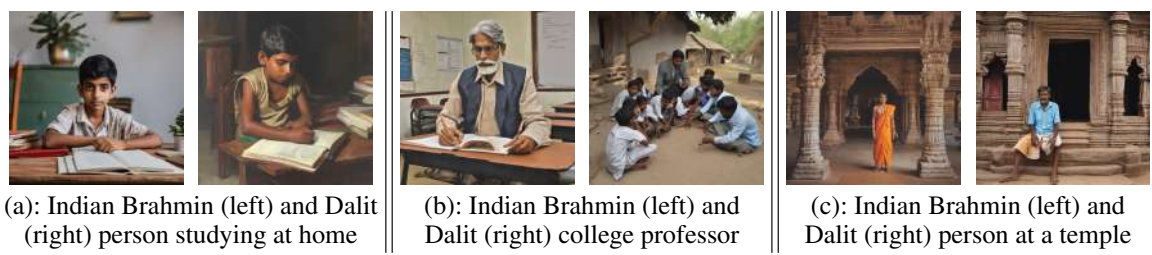

(a): Indian Brahmin (left) and Dalit (right) person studying at home

(b): Indian Brahmin (left) and Dalit (right) college professor

(c): Indian Brahmin (left) and Dalit (right) person at a temple for activities explored above), and explore T2I representations of that axis of identity within Indian-ness in performing various activities. As before, we generate 25 Stable Diffusion images and one ChatGPT/DALL-E3 image per activity.

Figure 8: Salient Evidence of Deeper Exoticism within T2I representations of Indian subpopulations, based on Caste.

The first aspect of identity under exploration is *caste*: an identity along which Indian society has subtly or overtly organized since the publication of the *Manusmriti* around 1 BC. Though believed to be absent within foundational texts of Hinduism from around 1500 BC, casteism has been encoded in Hindu society in India and functions as a rigid marker of privilege (or lack thereof). At a very high level, the modern caste system encodes four castes in descending order of religious privilege — Brahmins, Kshatriyas, Vaishyas, and Shudras — with a fifth group, the Dalits[2], placed outside the caste hierarchy and considered sub-human (see Ambedkar (1946) for a more detailed explanation). Dalit people have been victims of the social practice of untouchability, which legitimized and codified the act of considering groups of people as impure and were avoided by upper-caste people. Despite caste-based discrimination being illegal in India through the 1955 Protection of Civil Rights Act, the caste system is operationalized in Indian society through practices such as endogamy, segregation of housing, total/partial restrictions on Dalit people entering temples, restricting Dalits to manual labor and scavenging, and more. Caste abolitionists (e.g., Ambedkar 1948; Omvedt 2017; Soundararajan 2022; Wilkerson 2020; Yengde 2019) have long been advocating for the abolition of such practices, towards an Indian society without caste-based discrimination.

Casteist practices are also exacerbated by and encoded within technology. A Rudra, Urwin, and Calver (2020) investigation into the Indian matrimonial site Shaadi.com found casteist patterns where Brahmin users were almost exclusively shown Brahmin matches and very few matches with users from lower castes. NLP systems in English and Indian languages are also known to contain casteist tendencies by associating Dalits with negative sentiment (Dammu et al. 2024; Tiwari et al. 2022). Ghosh (2024a) and Ghosh (2024b)'s explorations into casteism within T2Is documented how representations of caste-oppressed groups were associated with menial labor and those for caste-privileged groups contained a sense of 'castelessness' where Savarna individuals claim to be casteless (and meritorious), while lower-caste individuals are still marked by caste (Vaghela, Jackson, and Sengers 2022). We extend their findings (they focused only on baseline 'Person of X caste' and 'Person of X caste at work') into a broader exploration on representations of caste-oppressed people, specifically Dalit people.

In our examination, clear patterns of casteism emerge within T2Is. Baseline representations of Indian people are known to paint rural and impoverished pictures (Ghosh et al. 2024; Qadri et al. 2023), and this trend is upheld when the word 'Dalit' is present within prompts, but not so for prompts containing the word 'Brahmin'. Consider Figure 8(a), which shows an Indian Brahmin (left) and Dalit (right) person studying at home. While both images depict boylike figures seated at tables reading books, a closer inspection of the setting might reveal that the Brahmin person is depicted in a nicer house with a chest of drawers and a plant on top, whereas the Dalit person is seated at a table that is quite small and low to the ground in an otherwise dull-appearing room. Figure 8(b) furthers this difference in economic prosperity, where an Indian college professor who is Brahmin (left) is shown seated inside what appears to a classroom but one who is Dalit (right) is outdoors with students seated on the ground, which also amplifies casteist tendencies of access to higher education. Beyond associations with poverty, casteist beliefs are also visible in Figure 8(c), where a Brahmin person (left) is shown inside a temple but the Dalit person (right) is shown outside, in line with casteist practices restricting the entry of Dalit people inside temples in India.

Beyond casteism, another aspect of identity along which extensive discrimination exists in India is *religion*, specifically against Muslim Indians. The landmass that is currently called India has a long history with the religion of Islam, dating back to the invasion of Mahmud of Ghazni in 997 AD. His military campaign established one of the first Islamic kingdoms in India and in 1025 AD, he raided and destroyed the famous Somnath Temple, one of India's largest temples and libraries at the time. Large parts of India were under Islamic rule through the Delhi Sultanate (1206–1526) and the Mughal Empire (1526 to 1857) right until colonization, with several instances of Hindu-Muslim kingdoms and armies fighting each other. Modern-day India was also partitioned twice along religious lines — once when Bengal was partitioned in 1905 and again when India and Pakistan were formed in 1947 as a result of independence from Britain — as religious tension was a crucial component of Britain's fa-

---
[2]The term 'Dalit' was coined by Indian social activist Jyotirao Phule from the Marathi word for those who were oppressed and broken by society. Dalits were also known as 'Harijans' ('children of God'), popularized by Mahatma Gandhi, but they found the term patronizing (Omvedt 2017), and instead largely prefer 'Dalit'.

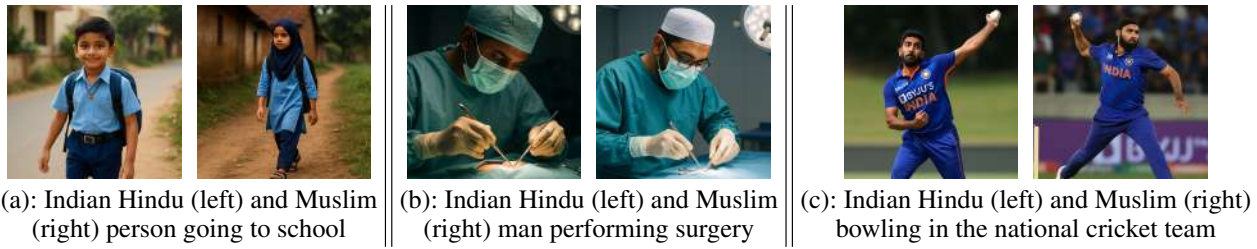

(a): Indian Hindu (left) and Muslim (right) person going to school

(b): Indian Hindu (left) and Muslim (right) man performing surgery

(c): Indian Hindu (left) and Muslim (right) bowling in the national cricket team mous 'Divide and Conquer' strategy in India. Communal violence has been prevalent in Indian history, including incidents such as the conflict over Kashmir in 1947-49, the 1983 Nellie Massacre in Assam, 1989 Bhagalpur Riots, the demolition of the Babri Masjid in 1990, the 2002 Gujarat riots, the 26/11 attack in 2008 Mumbai, and most recently, the Pahalgam terror attack of 2025 which sparked a skirmish between India and Pakistan, not to mention the three wars fought between India and Pakistan in 1965, 1971, and 1999.

Figure 9: Salient Evidence of Deeper Exoticism within T2I representations of Indian subpopulations, based on Religion.

Even though India is a secular nation, many in Indian politics believe that some of the policies of the current Indian government are Hindu-supremacist and anti-Muslim, driving the country towards a Hindu state. One such policy was the Citizenship Amendment Act (CAA) of 2019, which sought to provide Indian citizenship to refugees from neighboring countries with a specific focus on those facing religious persecution, thus extending only to non-Muslim refugees from Pakistan, Afghanistan, and Bangladesh. This was seen to be a targeted effort to increase the Hindu population in India, and sparked nationwide protests in 2019 and 2020. Today, even as 14.2% of the Indian population is Muslim (Chandramouli 2011), they are subtly or overtly subjected to discriminatory practices (e.g, Chakraborty and Bohara 2021; Krishnaswami 1978; Robinson 2008).

Within T2Is, Qadri et al. (2023) found that representations of Muslim people in Pakistan were confined to women wearing headscarves and men clad in prayer caps, within the larger context of 'Muslim == terrorist' identified in global contexts (e.g., Abrar et al. 2025; Bianchi et al. 2023). Our explorations of representations of Indian Muslims observed a similar trend. For instance, Figure 9(a) featuring representations of Indian people going to school rendered a boylike person when the prompt contained the word 'Hindu', but the output for the prompt containing 'Muslim' appears to depict a girllike figure in a headscarf, thus indicating how the mention of 'Muslim' invoked a change in gender to render a stereotypical association with the hijab. We see such exoticized representations continue in Figure 9(b), where Indian doctors performing surgery are both shown to cover their heads as is the sanitary requirement, but while the Hindu surgeon (left) uses the appropriate scrub-cap, the Muslim surgeon (right) is shown wearing the religious skull-cap. This is also true for Figure 9(c), when a Muslim bowler (right) playing for the Indian cricket team is also shown wearing a religious skull-cap while the Hindu bowler (right) is not, even though India has a long history of Muslim players representing the country and bowling without wearing skull-caps.

Finally, another aspect of identity along which subtle and/or overt discrimination is practiced in India is along *regional identity*, specifically against people from Northeast India. The Northeast of India consist of right States — Arunachal Pradesh, Assam, Meghalaya, Manipur, Mizoram, Nagaland, Sikkim, and Tripura — and has a long history of being Othered within India. During the early years of colonial rule in India, the British introduced the Inner Line Regulation, which separated parts of Northeast India from the rest of the country without a special permit. This was viewed as a way for the British to keep their people in 'mainland' India away from the Northeast, which predominantly consisted of tribal areas viewed as primitive (Gait 1906). In subsequent years, the region was treated as one of strategic importance because of its proximity to China, as Christian missionaries sought to gain their loyalties to the British Empire. This view of the Northeast as a strategic site continued post independence, as India integrated much of the region shortly following independence and also fought the Sino-Indian war of 1962 for land control of Arunachal Pradesh. The region has never been considered of much importance beyond this, as is reflected in historically low investments by the central government in its prosperity and its low representation in Indian industries, media, or mainstream sports. Due to residents of Northeast India having roots in East Asia and their Mongoloid features, they are often the subject of racist comments based on their appearances, described as 'barbaric' or rumored to eat dogs and snakes (Yadav and Badhe 2018). After the COVID-19 pandemic began and produced an uptick of racism against East Asian populations across the world, Northeastern Indians also faced similar issues, dealing with comments such as 'You people have brought the virus here' (EP, Negi, and AP 2022; Haokip 2021).

Explorations of bias against Northeast Indians is often absent in studies exploring bias against India, as such research merely cites their lack of coverage of the Northeast as a limitation (e.g., Ghosh et al. 2024). We begin our exploration with Figure 10(a), which depicts two Indian speakers at Universities. While the speaker from the northern Indian state of Delhi (left) is shown in an auditorium addressing a large group of people, the speaker from Nagaland (right) is shown not facing the speakers in a setting that appears outdoors. However, perhaps the most striking difference is in attire, whereby the speaker from Delhi and the students in the classroom are shown a wide range of colors, the speaker from Nagaland and others represented there are shown wearing red. This is a pattern we see continuing across other im-

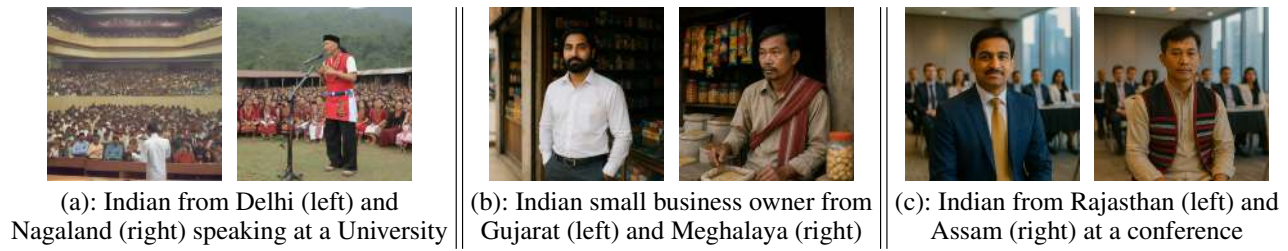

(a): Indian from Delhi (left) and Nagaland (right) speaking at a University

(b): Indian small business owner from Gujarat (left) and Meghalaya (right)

(c): Indian from Rajasthan (left) and Assam (right) at a conference ages: Figure 10(b) shows that DALLE-3 interprets 'Indian small business' as a grocery store and represents the owner of such a store from Gujarat as wearing a bright white shirt and pants, but the owner of a similar store in Meghalaya as a duller cream shirt with a red shawl; and Figure 10(c) shows that while conference attendees in India are all clad in Western formals, the attendee from Assam is shown in a similar dull cream shirt with a red-patterned sweater vest. Exoticism of Northeast India seems to be through the overrepresentation of red-colored attire, and it is not immediately clear to us why. An alarming reason would be because of the stereotypical association to China, but we do not have evidence for this. A more logical speculation is that red is possibly a prevalent color within traditional tribal attires from that region, though this team of researchers does have the appropriate positionality to definitively say so.

Figure 10: Salient Evidence of Deeper Exoticism within T2I representations of Indian subpopulations, based on Region.

We thus demonstrate patterns of exoticism within the outputs of T2Is ranging from exoticism of the "Global South", marginalized subpopulations of the "Global North" such as Indigenous people in the US and Australia, and deeper exoticism of marginalized subpopulations of the "Global South", i.e. caste-oppressed, Muslim, or Northeastern Indians.

## 5 Implications

We conclude with Implications towards multiple stakeholder groups within the GAI ecosystem, visualized in Figure 11.

### 5.1 For Everyday Users

**Be Aware of Exoticist Trends within T2Is:** A primary purpose of this paper is to make users from "Global South" countries and/or other populations historically marginalized in their respective contexts aware of the tendencies of T2Is like Stable Diffusion and DALLE-3 to possibly exoticize representations of their cultures. If non-use of such tools is not an option, we encourage users to rigorously engineer and re-engineer their prompts to include specificity, while also recognizing that this places an inequitable burden on marginalized users to work harder to avoid 'biased' outputs.

**Engage with GAI systems Critically:** Emerging technologies like GAI offer creative potential but risk distorting beauty standards even further, making it essential to help people critically navigate increasingly artificial representations. So we encourage users to engage critically with inclusive prompts using a guided framework of reflective questions to create more realistic and diverse outputs, and ask themselves guiding questions towards productive creative pursuits using AI: (i) Am I making the world a more inclusive place with my AI representation?, (ii) Do I have any preconceived notions about this group that would be helpful to check?, (iii) Who is going to feel represented by the image I generated? We take inspiration from the Dove commitment on Beauty in the AI age (Boechat and Diedrichs 2024), and instead of non-use, advocate for critical engagement.

**Value Cultural Literacy before Creativity:** If users are attempting in their prompts to generate images of individuals/communities they do not belong to, we encourage them to gather relevant cultural competency, either through secondary research or conversations with people with appropriate positionalities, to check generated images for exoticism. This is in line with human-centered research principles (e.g., Chancellor 2023) that cautions how creative expression through AI is powerful, but without cultural awareness, can easily reproduce stereotypes or exoticize communities.

**Reclaiming Marginalized Identities through GAI:** Finally, GAI tools can also be useful as agents of reclaiming narratives of marginalizations through positive, diverse representations and non-exoticized depictions of minoritized people in professional, joyful and spiritual settings. They can also be useful for imaginatively visualizing counterfactual histories and speculative futures, of worlds where "Global South" and marginalized populaitons were not oppressed through systemic structures of colonialism and imperialism.

### 5.2 For Researchers

**Stronger coverage for Global South contexts:** Based on our findings, we invite stronger coverage of "Global South" contexts similarly motivated as ours, in the spirit of expanding the common focus of AI 'bias' and fairness research which is often limited to gender (and occupation) and racial bias. We acknowledge, even based on our own research, that a large section of research on "Global South" contexts focuses on India and the subcontinent, as we invite researchers from other "Global South" populations to participate in broadening this coverage. Relatedly, we implore upon conferences such as AIES and FAccT, to put forward special tracks inviting such research and facilitating the participation of scholars from such populations through inclusive conference practices e.g., not requiring in-person presence (often a barrier for "Global South" attendees who have to navigate expensive and complicated visa and travel logistics to conferences typically held in "Global North" countries) for publication.

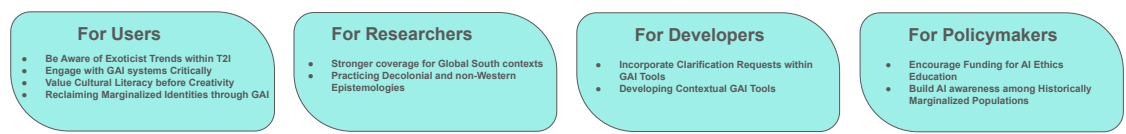

Figure 11: Visual Representation of the Implications of our work

**Practicing Decolonial and non-Western Epistemologies:** Given the imposition of the Western lens present within T2Is, we advocate for the inclusion of decolonial and non-Western epistemologies in the (re)design of such tools. For instance, while existing T2Is converge towards a 'most probable' output towards some objective representation of prompts, we encourage a pluralistic approach to produce a set of different outputs for users to choose (Lu and Van Kleek 2024). T2Is could also refuse to render 'default' versions of images requested through vague prompts, instead asking for further specificity before producing outputs. Notably, we do not provide extractivist recommendations such as 'seek stronger representation from Indigenous populations through data collection efforts' because while the idea of 'better data == better outputs' is not only akin to putting "a band aid solution" (D'ignazio and Klein 2020), but also fraught with danger given the histories of exploitation of marginalized people have suffered harm simply based on data collection efforts (Seltzer and Anderson 2001).

## 5.3 For Developers

**Incorporate Clarification Requests within GAI Tools:** Given that existing popular GAI tools are demonstrably prone to exoticizing marginalized populations, we encourage the developers of such tools to consider adding in features within their tools that seek clarification from users such that the outputs might be less prone to exoticism. For instance, given the prevalence of sombreros in association with Mexicans within GAI renderings, users including 'Mexican' in their prompts could be asked questions such as 'Text-to-image generators often associate representations of Mexican people with sombreros. Would you like to see sombreros in this set of outputs?' This is an overt way of requesting 'negative prompts', i.e., instructions on what to avoid within GAI outputs, that are not currently intuitive to include within the UIs of popular GAI tools.

**Developing Contextual GAI Tools:** Echoing Bender et al. (2021), we advocate for the development of context-specific ~~L~~LMs that are not intended to be used globally. We take inspiration from the machine translation tool Lesan (Hadgu, Aregawi, and Beaudoin 2022) designed to translate English to and from the Ethiopian languages of Tigrinya and Amharic, and the text-to-image generator Kalaido.ai (Lohchab 2024) designed to work in 17 Indian languages aside from English. The development of such contextually-specific tools, built using training data and with annotators from within a given context/culture, can allay the alarming outputs we provide above, as can be seen from Kalaido.ai's representation of Indian men and women (see Figure 14) performing the same activities explored in Section 2.

## 5.4 For Policymakers

**Encourage Funding for AI Ethics Education:** Given the popularity of GAI tools, there is a timely need to fund and introduce content about them in educational materials, as early as in middle school. It is essential for such curricula to include units about AI ethics and the ways in which existing GAI tools produce outputs harmful to marginalized populations. There have already been some units that have been developed, such as the Stanford Digital Education high school course on AI ethics (Rabinovitz 2025) and the STEAM in AI Data Science and AI Ethics College Prep Intensive course at the middle and high school levels that can be taken online by students across the US. Beyond AI harm in commonly explored areas such as gender or race bias, such curricula should also cover exoticism by GAI tools.

**Build AI awareness among Marginalized Populations:** We also encourage legislation around stronger protections for and awareness campaigns aimed at "Global South" and other historically marginalized populations. This is particularly critical for communities consuming GAI content without recognizing them as such, and making decisions based on information within such content. Consider, for instance, the 2024 Indian general elections where rural populations with historically low tech literacy were fed GAI-assisted campaign videos of candidates in their area directly speaking to them, which influenced how they voted (Raj 2024). A similar situation occurred in the African countries of Congo and South Africa, where GAI-created deepfakes were used to heavily influence public opinion around crucial topics (Allen and Nehring 2025). We urge policymakers with legislative authority over historically marginalized populations to better support their constituents through stronger GAI detection and awareness programs, as well as protect them from consuming GAI content without their consent.

## 6 Limitation, Future Work, and Conclusion

A limitation of our work is the usage of broad terms (e.g., 'Indigenous' or 'Dalit') to refer to large populations. While we fully recognize that this assimilates differences across vast subgroups into one term, our lived experiences do not allow us to accurately identify culturally-specific visuals unique to subpopulations, and we welcome closer scrutiny by individuals with more fine-grained lived experiences.

We only scratch the surface of the plethora of ways in which populations and subpopulations in the "Global South" are exoticized, as well as only explore explicit bias in T2Is. There remains a lot of scope for important future exploration of exoticism of marginalized populations within T2I outputs, and we hope to see such work in years to come.

# Adverse Impact and Ethical Considerations Statement

In doing AI ethics research as motivated as ours, it is essential that we reflect upon our own workflow and potential harm that can emerge from our research. Firstly, as alluded to earlier, our work contributes to a growing body of research (e.g., Ghosh et al. 2024; Qadri et al. 2023) that uses India as an example of "Global South" countries, and contributes towards a narrative of equating the two terms similarly as gender bias is often used as an all-encompassing example as bias against marginalized populations within NLP systems. While we do not go so far as to say that such a focus actively causes harm to other historically marginalized populations and subpopulations in the "Global South", we do acknowledge that the trend of using Indian focus as an interchangeable focus towards "Global South" contexts risks erasing several other such populations and subpopulations.

Additionally, similarly as Ghosh and Caliskan (2023b), we recognize that the very act of demonstrating harmful T2I depictions of marginalized populations within a publicly available paper or dataset can in fact contribute towards promoting more of such depictions in modern T2Is, if such image-caption pairs are scraped for model training. Therefore, upon acceptance of this paper and preparation of a public-facing version, we will replace the aforementioned images with blurred versions. We also do not make generated sets of images publicly available, and interested parties can reach out to the first author for controlled access.

## A  Model Specifications

All image generations in this study were conducted between March and May 2025 using Stability AI's stable-diffusion-xl-base-1.0 model[3], hosted on Hugging Face, and DALL-E 3, as integrated within the ChatGPT platform.

We note that large-scale generative models such as ChatGPT/DALL-E 3 and SDXL are subject to ongoing updates, including changes to their training data, model weights, inference behavior, and user interface. Accordingly, the outputs and behaviors analyzed in this study need to be understood in the context of the models' configurations during the data collection period. Future iterations of these models can yield different results under similar prompts and conditions. We therefore share all the obtained results through our narration.

## B  Prompt Details

To evaluate image generation models across geopolitical and cultural contexts, we created a structured set of text prompts that simulate everyday activities performed by individuals from various countries. These prompts follow a templated format to enable controlled comparisons.

**Prompt Template**

Each prompt was generated using the following template:

```
A [person / man / woman] from [COUNTRY] [ACTIVITY]
```

Example prompts:
- `A woman from Mexico playing soccer`
- `A person from the UK cooking`

We applied this template across 16 countries, 3 gender descriptors, and 10 everyday activities, resulting in 480 unique prompts. Each was run through the model twice (e.g., different seeds or model versions), totaling 960 generations.

**Countries Included**

- **Global South:** India, Bangladesh, Papua New Guinea, Egypt, Ethiopia, Tunisia, Sudan, Libya, Venezuela, Colombia, Indonesia, Honduras, Mexico
- **Global North (Baseline):** USA, UK, Australia

These countries were specifically chosen from the works: Ghosh et al. (2024); Venkit et al. (2023), that provided countries that were categorized as global north and south with high bias shown from a generative AI model.

**Activity Categories**

| Activity ID | Activity Description |
|---|---|
| 1 | hiking |
| 2 | flying |
| 3 | playing soccer |
| 4 | playing tennis |
| 5 | playing golf |
| 6 | cooking |
| 7 | at home |
| 8 | at work |
| 9 | grocery shopping |
| 10 | clubbing |

Table 1: List of activities used in prompt construction.

**Actor Variants**

**Sample Prompts**

**Prompt Generation Summary**

In total:
- 16 countries × 3 actor types × 10 activities = 480 prompts
- Each prompt repeated twice (e.g., for generation consistency): **960 total generations**

## C  Additional Results

---
[3]https://huggingface.co/stabilityai/stable-diffusion-xl-base-1.0

| Actor Type | Description |
|---|---|
| person | Gender-neutral subject |
| man | Masculine-coded subject |
| woman | Feminine-coded subject |

Table 2: Actor type variations in prompt formulation.

| Country | Actor Type | Prompt Example |
|---|---|---|
| India | person | A person from India hiking |
| Bangladesh | woman | A woman from Bangladesh cooking |
| USA | man | A man from the USA playing tennis |
| Colombia | person | A person from Colombia at home |
| UK | woman | A woman from the UK grocery shopping |
| Papua New Guinea | man | A man from Papua New Guinea clubbing |

Table 3: Example prompts demonstrating variation across country, actor type, and activity.

| Country | Person ~ Hiking | Person ~ Flight | Person ~ Cooking | Person ~ At Home | Person ~ Groceries | Person ~ Sports | Person ~ At Work |
|---|---|---|---|---|---|---|---|
| USA | 0.37 | 0.36 | 0.34 | 0.29 | 0.33 | 0.36 | 0.48 |
| UK | 0.41 | 0.43 | 0.44 | 0.38 | 0.42 | 0.42 | 0.41 |
| Australia | 0.44 | 0.45 | 0.52 | 0.37 | 0.32 | 0.44 | 0.41 |
| India | 0.76 | 0.68 | 0.74 | 0.76 | 0.64 | 0.61 | 0.71 |
| Bangladesh | 0.74 | 0.69 | 0.78 | 0.66 | 0.67 | 0.64 | 0.77 |
| Indonesia | 0.81 | 0.73 | 0.73 | 0.76 | 0.74 | 0.71 | 0.81 |
| Ethiopia | 0.71 | 0.59 | 0.69 | 0.76 | 0.64 | 0.63 | 0.77 |
| Colombia | 0.68 | 0.46 | 0.68 | 0.69 | 0.79 | 0.51 | 0.58 |
| Papua New Guinea | 0.91 | 0.89 | 0.91 | 0.93 | 0.86 | 0.73 | 0.90 |
| Egypt | 0.70 | 0.73 | 0.72 | 0.81 | 0.73 | 0.68 | 0.76 |
| Tunisia | 0.72 | 0.74 | 0.74 | 0.80 | 0.74 | 0.68 | 0.79 |
| Sudan | 0.92 | 0.77 | 0.72 | 0.91 | 0.76 | 0.71 | 0.89 |
| Libya | 0.73 | 0.71 | 0.81 | 0.91 | 0.74 | 0.63 | 0.83 |
| Venezuela | 0.69 | 0.68 | 0.71 | 0.83 | 0.71 | 0.66 | 0.63 |
| Honduras | 0.52 | 0.62 | 0.73 | 0.88 | 0.69 | 0.67 | 0.68 |
| Mexico | 0.76 | 0.59 | 0.76 | 0.87 | 0.77 | 0.74 | 0.71 |

Table 4: Cosine Similarity scores (ranging 0-1) for comparisons **by country**. For a given country, each score is computed by comparing the baseline 'Person' images for that country with the respective activity *e.g.,* comparing the images for 'Person from the USA' with 'Person from the USA on a hike' results in an average cosine similarly score of 0.47. A lower score is indicative of a larger difference across images which is interpreted as a wider variation of attires and displaying activity-appropriate attires whereas a higher score is interpreted as the presence of consistent and stereotypical outfits irrespective of activity. The results for 'Sports' column is an average across the comparison scores for three sports (Soccer, Tennis, and Golf).

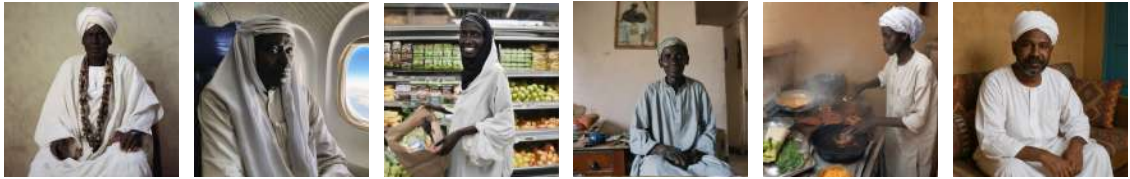

Figure 12: Salient T2I Outputs showing Exoticized depictions of Sudanese people, with the persistence of white flowing robes and turbans. The first image is a baseline Sudanese person, and subsequent images depict Sudanese people performing the activities (from left to right) taking a flight, getting groceries, being at home, cooking, and being at home. The first 5 images are Stable Diffusion generated (one of 25 images shown), and the last one is by ChatGPT/DALL-E 3.

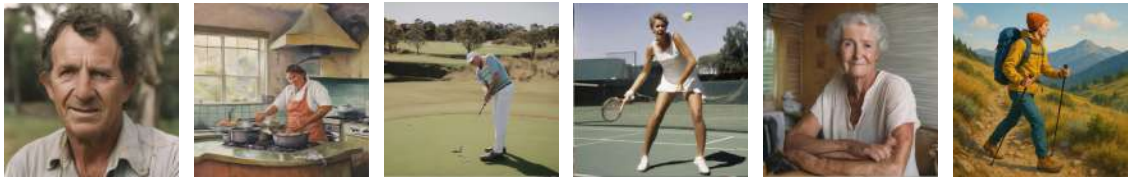

Figure 13: Salient T2I Outputs showing non-Exoticized depictions of Australian people, who are shown in activity-appropriate attire for the activities of (from left to right) cooking, playing golf, playing tennis, being at home, hiking, and hiking. The first 5 images are Stable Diffusion generated (one of 25 images shown), and the last one is by ChatGPT/DALL-E 3.

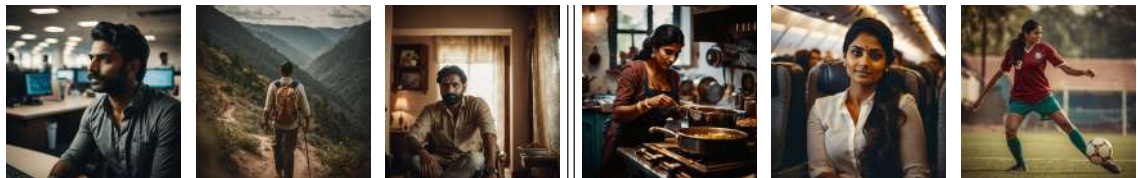

Indian man at home (left), on a hike (center), and at home (right)

Indian woman cooking (left), on a flight (center), and playing soccer (right)

Figure 14: Non-Exoticized representations of Indian men and women performing various activities, as produced by the Indian text-to-image generator Kalaido.ai.